\documentclass{article}
\usepackage{amsmath}
\usepackage{graphicx}
\usepackage{amsfonts}
\usepackage{amssymb}
\usepackage{geometry}
\usepackage{float}
\usepackage{bm}
\usepackage{color}
\usepackage{cite}

\begin{document}

\title{New higher-order generalized uncertainty principle: Applications}
\author{B. Hamil \\
D\'{e}partement de TC de SNV, Universit\'{e} Hassiba Benbouali, Chlef,
Algeria. \and B. C. L\"{u}tf\"{u}o\u{g}lu \\
Department of Physics, Akdeniz University, Campus 07058, Antalya, Turkey, \\
Department of Physics, University of Hradec Kr\'{a}lov\'{e}, \\
Rokitansk\'{e}%
ho 62, 500 03 Hradec Kr\'{a}lov\'{e}, Czechia.}
\date{}
\maketitle

\begin{abstract}
Last year, Chung and Hassanabadi proposed a higher order general uncertainty
principle (GUP$^\ast$) that predicts a minimal length as well as possesses a
upper bound momentum limit. In this article, we have discussed an ideal gas
system and its thermal properties using that deformed canonical algebra
introduced by them. Moreover, we examined blackbody radiation spectrum and
the cosmological constant in the presence of the GUP$^\ast$. After a
comparison with the existing literature, we concluded that the given
formalism of Chung and Hassanabadi yields more accurate results.
\end{abstract}

\begin{description}
\item[PACS numbers:] 04.60.-m, 04.60.Bc, 05.90.+m.

\item[Keywords:] A new higher order GUP, Ideal Gas, Blackbody, the
Cosmological constant.

\item[E-mail addresses:] hamilbilel@gmail.com
(B.Hamil),bclutfuoglu@akdeniz.edu.tr (B. C. L\"{u}tf\"{u}o\u{g}lu).
\end{description}

\section{Introduction}

In quantum mechanics, there is the concept of wave-particle duality, which
has no counterpart in classical mechanics. According to this concept, it is
argued that every quantum quantity has a particle and wave property. For
some pairs of physical observables of a pointlike considered particle, such
as its position and momentum, a fundamental limit exists on the measurement
of them simultaneously, which is governed with the Heisenberg uncertainty
principle (HUP). However, the uncertainty values of any of these observables
can be arbitrarily small. Contrarily, researches based on the string theory 
\cite{Veneziaono1986, Amatietal1989, Amatietal1990, Konishietal1990,
Amatietal1993, Witten1996}, background indepedendent \cite{Garay1995,
Dasetal2008} and dependent \cite{Susskind, Weet, Hull, Balasub1, Witten,
Strominger1, Strominger2, Balasub2} quantum theory of gravity, doubly
special relativity (DSR) theories \cite{magueijo1, magueijo2, Cortesetal2005}%
, non commutative space-time \cite{Maggioe1993, Kempef, Kempf1996,
Kempfetal997} and field theory \cite{Douglas}, black hole physics \cite%
{Scardiglietal2003, Ovgun2016, Ovgunetal2016, Ovgunetal2017, H1, H2, H3, H4,
H5, H6, H7}, and gedanken experiments \cite{Maggioe1993GE, Scardigli1999,
Tawfik}, etc. predict a finite lower limit value for these measurements \cite%
{Tawfik, Tawfiketal2015}. This lower bound is called the minimum length (ML)
and it is assumed to be on the Planck scale. In this context, an ML can be
regarded as the fuzziness of space-time \cite{Kempf1997}. Technically, in
order to derive an ML scale out of the canonical commutation relations
(CCR), some small correction terms are implemented \cite{Kempf1994}. Thus,
the HUP is modified to a more general form of it, namely the generalized
uncertainty principle (GUP).

Kempf \emph{et al.}, in \cite{Kempf2001, Kempfetal2001}, showed that the
formulation of the generalization can not be uniquely expressed \cite%
{Changetal20022}. For instance, in one approach \cite{Kempef}, Kempf with
Mangano and Mann (KMM) deformed the CCR in one dimension as 
\begin{equation}
\lbrack X,P]=i\hbar (1+\beta P^{2}),
\end{equation}%
where $\beta $ is the deformation parameter and it is $\beta \sim 10^{19}$ $%
\text{GeV}$. Then, they obtained the GUP in the form of 
\begin{equation}
\left( \Delta X\right) \left( \Delta P\right) \geq \frac{\hbar }{2}\left[
1+\beta ^{2}\left( \Delta P\right) ^{2}+\beta \langle P^{2}\rangle \right] ,
\end{equation}%
and derived the ML in positions as $(\Delta x_{0})_{KMM}=\hbar \sqrt{\beta }$%
. However, the lower bound limit in position does not predict a maximal
observable momentum value \cite{Nozarietal2012}. To embody an upper bound
limit to the observable momentum, Nouicer \cite{Nouicer2007} and Ali \emph{%
et al.} \cite{Alietal2009, Dasetal2010, Alietal2011} proposed two different
deformation to the CCR. Within the perturbative approximation, they obtained
uncertainty in position and momentum proportional and inversely proportional
to the deformation parameter, respectively. It is worth noting that these
results are valid only for small deformation parameters. Moreover, the
findings are not appropriate to the DSR theories since maximal momentum
differs from the maximal momentum uncertainty \cite{magueijo1, magueijo2,
Cortesetal2005, Pedram2012}. To overcome these objections, a higher-order of
GUP (GUP$^{\ast }$) is considered \cite{pedram, Pedram2010, Pedrametal2011,
Changetal20021, Shababia2017, chung}. For example, Pedram, in 2012, proposed
a higher-order GUP$^{\ast }$ \cite{Pedram2012, pedram} with the following
deformed CCR 
\begin{equation}
\lbrack X,P]=\frac{i\hbar }{1-\beta ^{2}P^{2}}.
\end{equation}%
Since the deformed CCR is singular at a certain momentum value, i.e. $%
P=1/\beta $, the particle's momentum can not exceed a certain maximal value.
Thus, the objections due to the DSR theories are eliminated. In this
higher-order GUP scenario, the smallest observable uncertainty in positions
reduces to $(\Delta x_{0})_{Pedram}=\frac{3\sqrt{3}}{4}\hbar \sqrt{\beta }$.
In a comparison with other scenarios, Pedram showed that $(\Delta
x_{0})_{KMM}<(\Delta x_{0})_{Noucier}<(\Delta x_{0})_{Pedram}$. Moreover, he
extended the GUP formalism to $D$-dimension and discussed several
applications such as hydrogen atom, harmonic oscillator and free particle's
solutions, a confined particle in a box, cosmological constant, and black
body radiation etc.. \cite{pedram, Pedram2012, Pedram2013}. However, in this
formalism, the energy spectrum of the position operator's eigenfunctions has
a divergency problem.

Very recently, Chung and Hassanabadi proposed a new higher-order GUP (GUP$%
^{\ast }$) formalism \cite{chung} that has an upper bound momentum value in
the following form: 
\begin{equation}
\left[ X,P\right] =\frac{i\hbar }{1-\alpha \left \vert P\right \vert },
\quad \alpha>0,  \label{1a}
\end{equation}
where $\left \vert P\right \vert= \sqrt{\left \vert P^2\right \vert}$. They
examined the algebraic structure of the GUP$^{\ast }$ formalism and showed
that it is different than Pedram's approach. After a detailed examination
they concluded that in the latter scenario the divergency problem of the
Pedram's formalism does not arise \cite{chung}.

In this paper, we are motivated to investigate a classical ideal gas system
within a statistical mechanics context by means of the thermodynamic
functions according to the GUP$^{\ast }$. Furthermore, we revisit the
blackbody radiation and cosmological constant problems according to the new
framework. We construct the paper as follows: In the next section, we review
the GUP$^{\ast }$. In section \ref{IG}, we discuss how to construct a
partition function with the deformed formalism. Then, we consider an ideal
gas that is constituted from a monatomic atom in a canonical ensemble and
derive the thermodynamic functions. We analyze the results with numerical
values. In section \ref{App}, we employ the new formalism in two other
quantum phenomena, namely the blackbody radiation and cosmological constant
problems. Finally we end our paper, with a brief conclusion section.

\section{A new higher order GUP}

In this contribution, we use the GUP$^{\ast }$ formalism which is given by
Chung and Hassanabadi \cite{chung}. In one dimension, they presented the
deformed CCR between the position $X$ and momentum $P$ operators as written
in Eq. \eqref{1a}. Here, the deformation parameter 
$\alpha =\alpha _{0}/\left( m_{P}c\right) ,$ where $m_{P}$ is the Planck
mass and $\alpha _{0}$ is of the order of the unity. It is worth noting
that, alike the Pedram's work, the CCR contains a singularity at $\left \vert
P\right \vert =1/\alpha $, which means that the momentum of the particle
cannot surpass $1/\alpha $, thus, it is compatible with the DSR theory \cite%
{magueijo1, magueijo2, Cortesetal2005, Pedram2012, chung}. Furthermore, the
physical observables such as momentum and the energy are not only
nonsingular, but also they are bounded from above. The uncertainty relation
that appears in the GUP$^{\ast }$ approach is found in the form of 
\begin{eqnarray}
\left( \Delta X\right) \left( \Delta P\right)  &\geq &\frac{\hbar }{2}%
\left \langle \frac{1}{1-\alpha \left \vert P\right \vert }\right \rangle , 
\notag \\
&=&\frac{\hbar }{2}\left[ \frac{1}{1-\alpha \left( \Delta P\right) }-\alpha
\left( \Delta P\right) \right] .
\end{eqnarray}%
For $\left( \Delta P\right) =1/(2\alpha )$, the ML uncertainty reaches its
minimum value 
\begin{equation}
\left( \Delta X\right) _{\min }=\frac{3\hbar \alpha }{2}.
\end{equation}%
In the momentum space we employ the following realization of the position
and momentum operators 
\begin{equation}
X=\frac{i\hbar }{1-\alpha \left \vert p\right \vert }\frac{d}{dp};\  \ P=p.
\end{equation}%
which satisfy the deformed CRR given in Eq. (\ref{1a}). In this case the
scalar product is not the usual one, instead it is defined with%
\begin{equation}
\left \langle \phi \right \vert \left. \psi \right \rangle =\int_{-1/\alpha
}^{1/\alpha }dp\left( 1-\alpha \left \vert p\right \vert \right) \phi ^{\ast
}\left( p\right) \psi \left( p\right) .
\end{equation}%
This definition preserve the hermiticity of the position and momentum
operators. It is worth noting that the measure of the momentum space is
modified, i.e.,%
\begin{equation}
dp\rightarrow dp\left( 1-\alpha \left \vert p\right \vert \right) .
\end{equation}%
In the classical domain, the commutator in quantum mechanics is replaced
poisson bracket as, 
\begin{equation}
\frac{1}{i\hbar }\left[ X,P\right] \rightarrow \left \{ X,P\right \} .
\end{equation}%
Consequently the classical limit of Eq. (\ref{1a}) give%
\begin{equation}
\left \{ X,P\right \} =\frac{1}{1-\alpha \left \vert P\right \vert }.
\end{equation}

\section{Ideal gas and its thermal quantities}

\label{IG} In this section, in the framework of the GUP$^{\ast }$ we
investigate the thermal quantities of an ideal gas. In this context, we
first discuss the construction of a partition function in the deformed
algebra. Then, we assume that an ideal gas in canonical ensemble and examine
its thermal quantities by obtaining the Helmholtz free energy, internal
energy, entropy and specific heat functions. Finally, we assign numerical
values and present the effect of the deformation parameter on the thermal
quantities.

\subsection{Construction of the partition function}

There are two possibilities to establish a partition function of a single
particle in a canonical ensemble in the presence of the GUP$^{\ast }$ \cite%
{Nozari,Nozari2,Vakili,Kamali,Fityo,pedram}:

\begin{enumerate}
\item Taking into account the commutation relations in the presence of the
GUP$^{\ast }$ which are simultaneous to the undeformed Hamiltonian.

\item Employing the standard commutation relation, however considering the
deformed Hamiltonian.
\end{enumerate}

In this paper, we prefer to use the first method. Therefore, for a system
with one particles, we construct the partition function in the presence of
GUP$^{\ast }$ with 
\begin{equation}
Z=\frac{1}{h^{3}}\int \int \frac{d^{3}Xd^{3}P}{J}e^{-\beta H},
\end{equation}%
where $\beta =\left( K_{B}T\right) ^{-1},$ $K_{B}$ is the Boltzmann
constant. Furthermore, $T$ represents the thermodynamic temperature and $J$
is the Jacobian of the transformation 
\begin{equation}
J=\prod_{j=1}^{3}\left \{ X_{j},P_{j}\right \} =\left( 1-\alpha \left \vert
P\right \vert \right) ^{3}.  \label{2b}
\end{equation}
The Hamiltonian of classical ideal gas is only a function of momenta, i.e. $%
H=\frac{P^{2}}{2m}$. For simplicity, we work in one dimension, thus, the
partition function for a single particle reads as 
\begin{equation}
Z=\frac{1}{h}\int \int \frac{dXdP}{J}e^{-\beta H}=\frac{L}{h}\int_{-1/\alpha
}^{1/\alpha }\left( 1-\alpha \left \vert P\right \vert \right) e^{-\frac{%
\beta P^{2}}{2m}}dP.  \label{A}
\end{equation}%
The integral has an exact solution. After simple algebra, we get 
\begin{equation}
Z=\frac{L}{\hbar }\sqrt{\frac{m}{2\pi \beta }}\left[ \mathrm{erf}\left( 
\sqrt{\frac{\beta }{2\alpha ^{2}m}}\right) -2\alpha \sqrt{\frac{m}{2\pi
\beta }}\left( 1-e^{-\frac{\beta }{2\alpha ^{2}m}}\right) \right] ,
\label{B}
\end{equation}%
where $\mathrm{erf}\left( x\right) $ is the "error function". As $\alpha $
is supposed to be a small parameter, one can expand Eq. (\ref{B}) up to
first order of $\alpha $, this yields 
\begin{equation}
Z\simeq \frac{L}{\hbar }\sqrt{\frac{m}{2\pi \beta }}\left[ 1-2\alpha \sqrt{%
\frac{m}{2\pi \beta }}\right] .  \label{Ba}
\end{equation}
The first term of Eq. \eqref{Ba} is the conventional partition function of
the ideal gas, while the second term represents the quantum gravity
correction.

\subsection{Thermal quantities}

In this subsection, based on the obtained partition function we derive some
thermodynamic functions. We start with the Helmholtz free energy,%
\begin{equation}
F=-\frac{1}{\beta }\ln Z=-\frac{1}{\beta }\ln \frac{L}{\hbar }\sqrt{\frac{m}{%
2\pi \beta }}-\frac{1}{\beta }\ln \left[ \mathrm{erf}\left( \sqrt{\frac{%
\beta }{2\alpha ^{2}m}}\right) -2\alpha \sqrt{\frac{m}{2\pi \beta }}\left(
1-e^{-\frac{\beta }{2\alpha ^{2}m}}\right) \right] .  \label{C}
\end{equation}%
Analogously, for very small deformation parameter the Helmholtz free energy
reduces to the following form%
\begin{equation}
F\simeq -\frac{1}{\beta }\ln \frac{L}{\hbar }\sqrt{\frac{m}{2\pi \beta }}+%
\frac{2\alpha }{\beta }\sqrt{\frac{m}{2\pi \beta }}.  \label{Helm1}
\end{equation}%
Here, the first term is the ordinary Helmholtz free energy for the one
dimensional classical ideal gas and the second term is the modification due
to the GUP$^{\ast }$ effects. For the internal energy, we utilize the
following formula%
\begin{equation}
U=-\frac{\partial }{\partial \beta }\ln Z.  \label{D}
\end{equation}%
Inserting (\ref{B}) into Eq.(\ref{D}) yields:%
\begin{equation}
U=\frac{1}{2\beta }-\frac{\frac{\alpha ^{2}m}{\beta }}{\sqrt{2\pi m\alpha
^{2}\beta }}\frac{1-e^{-\frac{\beta }{2m\alpha ^{2}}}}{\left[ \mathrm{erf}%
\left( \sqrt{\frac{\beta }{2\alpha ^{2}m}}\right) -2\alpha \sqrt{\frac{m}{%
2\pi \beta }}\left( 1-e^{-\frac{\beta }{2\alpha ^{2}m}}\right) \right] }.
\label{E}
\end{equation}%
In the limit of $\alpha \ll 1$, up to the first order of $\alpha $ we arrive
at 
\begin{equation}
U\simeq \frac{1}{2\beta }-\frac{\alpha }{\beta }\sqrt{\frac{m}{2\pi \beta }}.
\label{IntE1}
\end{equation}%
Next, we take the mass of the monatomic particle, $m=10^{-27}$~$kg$, as the
ideal gas system into consideration and we calculate the ratio of the
internal energy shift to the internal energy. 
\begin{equation}
\frac{\Delta U}{U_{\alpha =0}}\simeq \alpha \left( K_{B}T\right) ^{\frac{1}{2%
}}\times 10^{-14}.
\end{equation}%
We see that the GUP$^{\ast }$ induced correction term is very small to be
measured in low temperature. Therefore, we conclude that the quantum gravity
effect modifies the thermal quantities only at very high temperature limits.
Then, we study another important thermodynamic function, namely the entropy.
We use the well-known definition of the reduced entropy function 
\begin{equation}
\frac{S}{K_{B}}=\beta ^{2}\frac{\partial }{\partial \beta }F.  \label{F}
\end{equation}%
Substituting Eq. (\ref{C}) into Eq. (\ref{F}), we find%
\begin{eqnarray}
\frac{S}{K_{B}} &=&\frac{1}{2}+\ln \frac{L}{\hbar }\sqrt{\frac{m}{2\pi \beta 
}}-\frac{\frac{\alpha m}{\sqrt{2\pi m\beta }}\left( 1-e^{-\frac{\beta }{%
2m\alpha ^{2}}}\right) }{\left[ \mathrm{erf}\left( \sqrt{\frac{\beta }{%
2\alpha ^{2}m}}\right) -2\alpha \sqrt{\frac{m}{2\pi \beta }}\left( 1-e^{-%
\frac{\beta }{2\alpha ^{2}m}}\right) \right] }  \notag \\
&+&\ln \left[ \mathrm{erf}\left( \sqrt{\frac{\beta }{2\alpha ^{2}m}}\right)
-2\alpha \sqrt{\frac{m}{2\pi \beta }}\left( 1-e^{-\frac{\beta }{2\alpha ^{2}m%
}}\right) \right] .
\end{eqnarray}%
In the limit where $\alpha \ll 1,$ one finds the reduced entropy function up
to the first order of the deformation parameter as%
\begin{equation}
\frac{S}{K_{B}}\simeq \frac{1}{2}+\ln \frac{L}{\hbar }\sqrt{\frac{m}{2\pi
\beta }}-\alpha \sqrt{\frac{m}{2\pi \beta }}.  \label{Ent1}
\end{equation}%
We see that the GUP$^{\ast }$ ends with a negative correction term, which
conducts to a decrease on the reduced entropy of the system. It is
noteworthy that for $\alpha =0,$ the ordinary entropy is obtained. Finally,
we explore the reduced specific heat function of the system by employing the
following formula:%
\begin{equation}
\frac{C}{K_{B}}=-\beta ^{2}\frac{\partial U}{\partial \beta }.  \label{I}
\end{equation}%
By substituting Eq. (\ref{E}) into Eq. (\ref{I}), we arrive at%
\begin{eqnarray}
\frac{C}{K_{B}} &=&\frac{1}{2}-\frac{3}{2}\frac{\alpha m}{\sqrt{2\pi m\beta }%
}\frac{1-e^{-\frac{\beta }{2m\alpha ^{2}}}}{\left[ \mathrm{erf}\left( \sqrt{%
\frac{\beta }{2\alpha ^{2}m}}\right) -2\alpha \sqrt{\frac{m}{2\pi \beta }}%
\left( 1-e^{-\frac{\beta }{2\alpha ^{2}m}}\right) \right] }  \notag \\
&&+\frac{\alpha m\beta }{\sqrt{2\pi m\beta }}\frac{\frac{1}{2m\alpha ^{2}}%
e^{-\frac{\beta }{2m\alpha ^{2}}}\left[ \mathrm{erf}\left( \sqrt{\frac{\beta 
}{2\alpha ^{2}m}}\right) -2\alpha \sqrt{\frac{m}{2\pi \beta }}\left( 1-e^{-%
\frac{\beta }{2\alpha ^{2}m}}\right) \right] -\frac{m\alpha }{\beta \sqrt{%
2\pi m\beta }}\left( 1-e^{-\frac{\beta }{2m\alpha ^{2}}}\right) ^{2}}{\left[ 
\mathrm{erf}\left( \sqrt{\frac{\beta }{2\alpha ^{2}m}}\right) -2\alpha \sqrt{%
\frac{m}{2\pi \beta }}\left( 1-e^{-\frac{\beta }{2\alpha ^{2}m}}\right) %
\right] ^{2}}.  \label{G}
\end{eqnarray}%
Considering a very small deformation parameter, we expand Eq. (\ref{G}) up
to the first order in $\alpha $. We find 
\begin{equation}
\frac{C}{K_{B}}\simeq \frac{1}{2}-\frac{3\alpha }{2}\sqrt{\frac{m}{2\pi
\beta }}.  \label{Spec1}
\end{equation}%
The first term, which has a constant value, is the conventional reduced
specific heat function of an ideal gas in one dimension. The second term is
the correction term. In order to predict at which temperature the correction
term would have a significant effect, we examine the ratio of the reduced
specific shift to the reduced specific heat function. 
\begin{equation}
\frac{\Delta C}{C_{\alpha =0}}=3\alpha \sqrt{\frac{m}{2\pi \beta }},
\end{equation}%
For a hydrogen atom, which has a mass $m=9.11\times 10^{-31}kg$, at room
temperature $T=300K$, the evaluation of the ratio gives %
\begin{equation}
\frac{\Delta C}{C_{\alpha =0}}\simeq \alpha \times 7.\, \allowbreak
351\,8\times 10^{-26}.
\end{equation}%
The accuracy concerning the measurement of the specific heat is about $%
10^{-7}$ \cite{Mohr}$.$ Considering this precision, we get the following
upper bound for the minimal length as: :%
\begin{equation}
\left( \Delta X\right) _{\min }\simeq 0.2\, \allowbreak 056\,4\text{ \textrm{%
fm}}.
\end{equation}%
This bound is consistent with \cite{Nouicer,Brau,Akhoury}.

\subsubsection{Numerical Results and Discussions}

In this subsection, we use numerical values to examine our findings
graphically. We consider a non-interacting monatomic ideal gas. We take $%
m=10^{-27}$~$kg$, $L=1$~$m$ and use the Boltzmann and the reduced Planck
constants to plot the thermodynamic functions. In all graphs, we employ
three non-zero deformation parameters to demonstrate the effect of the GUP$%
^{\ast }$. In the figures, we employ the black and solid lines for
presenting the non deformed case.

At first, we plot the Helmholtz free energy versus temperature in Fig. \ref%
{fig1}. We see that at low temperatures the GUP$^{\ast }$ is not observable.
In very high temperatures, the Helmholtz free energy function has a shift
upwards. This result is in a complete agreement with Eq. \eqref{Helm1}. For
higher values of the deformation parameter, the shift becomes larger. 
\begin{figure}[hbtp]
\centering
\includegraphics[scale=1]{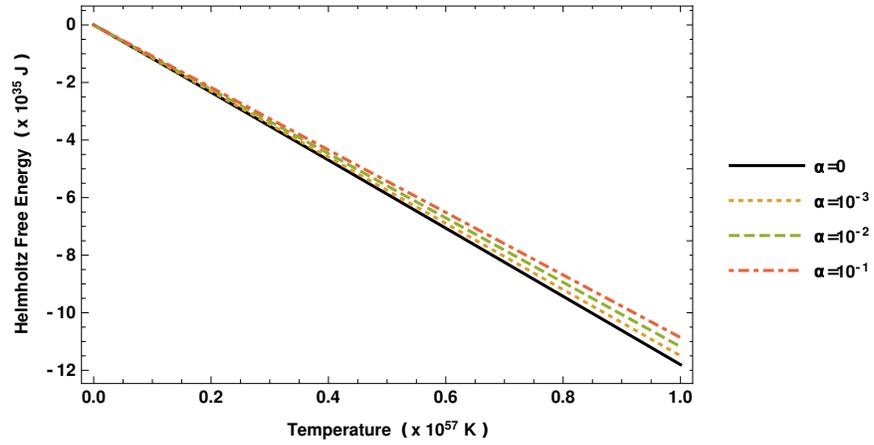}  
\caption{Helmholtz free energy versus temperature.}
\label{fig1}
\end{figure}

\begin{figure}[hbtp]
\centering
\includegraphics[scale=1]{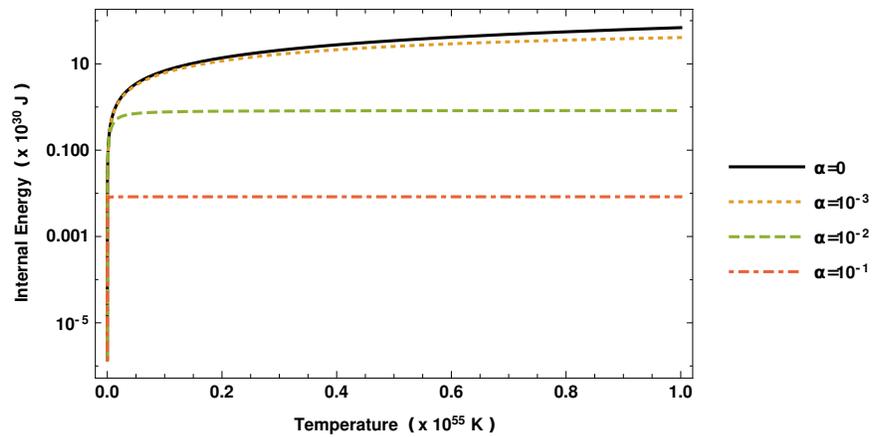}  
\caption{Internal energy versus temperature.}
\label{fig2}
\end{figure}

We schematically depict the behavior of the internal energy function versus
temperature in Fig. \ref{fig2}. At low temperatures, the internal energy has
a linear increase and the effect of the GUP$^{\ast }$ does not arise.
However at very high temperatures, the internal energy function takes lower
values with the higher deformation parameter. This behavior confirms Eq. %
\eqref{IntE1}.  

\begin{figure}[hbtp]
\centering
\includegraphics[scale=1]{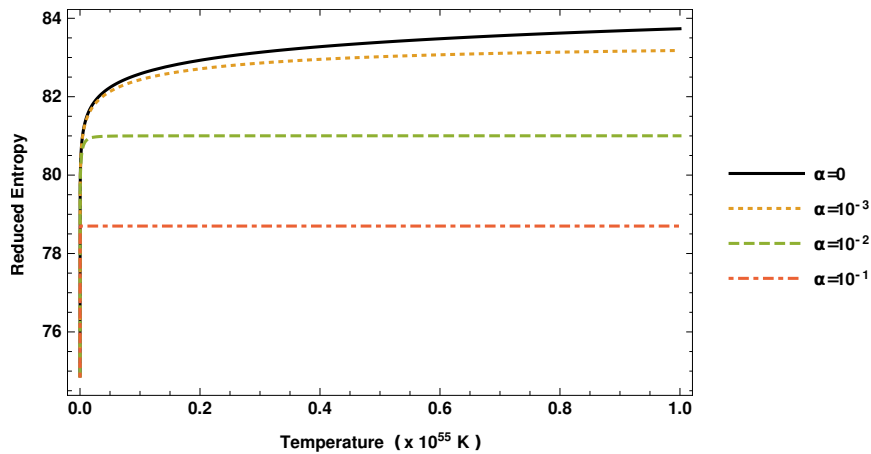}  
\caption{Reduced entropy function versus temperature.}
\label{fig3}
\end{figure}

Next, we present the behavior of the reduced entropy function versus
temperature in Fig. \ref{fig3}. We do not observe the effect of the GUP$%
^{\ast }$ at low temperatures, therefore, in the plot, we do not focus on
that region. At very high temperatures the entropy function has smaller
values via the higher values of the deformation parameter. Eq. \eqref{Ent1}
has a negative valued expression which is linearly proportional to the
deformation parameter, so that, we conclude that the graphical
representations of the reduced entropy are in an agreement with our
findings.  

\begin{figure}[hbtp]
\centering
\includegraphics[scale=1]{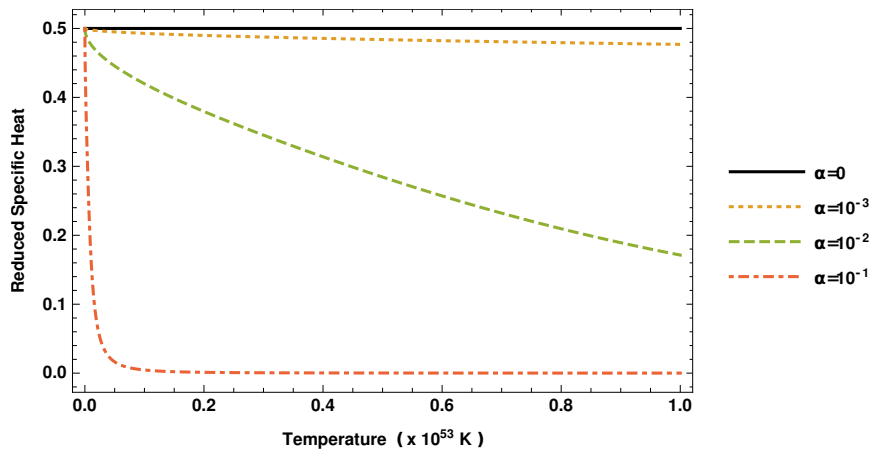}  
\caption{Reduced specific heat function versus temperature.}
\label{fig4}
\end{figure}

Finally, we plot the reduced specific heat function versus temperature in
Fig. \ref{fig4}. We see that in the ordinary case the specific heat function
is a constant. In the GUP$^{\ast }$ framework at low temperatures we do not
observe a modification. However, in very high temperatures we see a decrease
in the reduced specific heat function. These changes differ via the values
of the deformation parameter. Eq. \eqref{Spec1} approves such behaviors.  

All these figures prove that the thermal properties of the ideal gas are
affected by the deformation in the GUP$^{\ast }$ approach. We observe the
effects become significant only in the very high-temperatures. It is worth
noting that at low-temperature values the thermal properties are the same,
thus, they do not observed.

\section{Other applications}

\label{App} In this paper, we consider two more applications which are
frequently discussed in the contents of the works of deformed algebra: The
blackbody radiation spectrum and the cosmological constant.

\subsection{The blackbody radiation spectrum}

\begin{figure}[hbtp]
\centering
\includegraphics[scale=1]{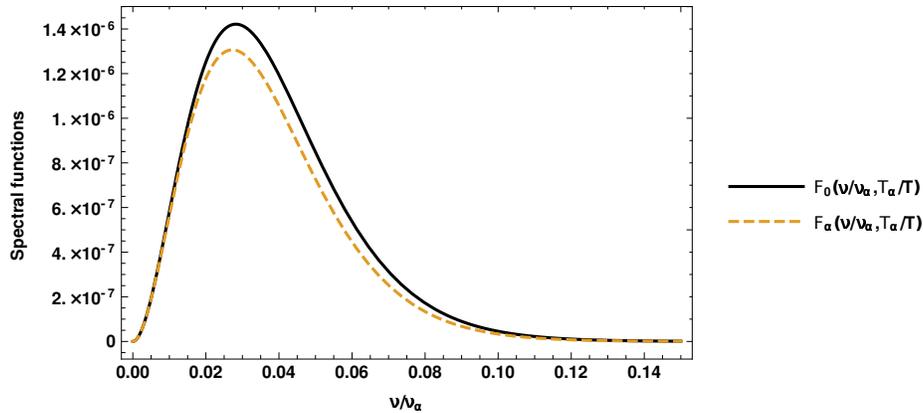}  
\caption{The blackbody radiation spectrum versus frequency in the GUP$^\ast$
framework at temperature $T=0.01T_{\protect \alpha }$.}
\label{fig5}
\end{figure}

In this subsection, we investigate the influence of GUP$^{\ast }$ on the
spectrum of the black body radiation. According to Eq. (\ref{2b}), the
number of quantum states per momentum space volume is modified by the weight
factor that is given in 3-dimensions as $J=1/\left( 1-\alpha \left \vert
P\right \vert \right) ^{3}$. So that, we express the energy density of the
electromagnetic field per unit volume at a finite temperature by \cite%
{Changetal20022, Pedram2012} 
\begin{eqnarray}
E &=&2\int \frac{d^{3}k}{\left( 2\pi \right) ^{3}}\left( 1-\alpha \hbar
\left \vert k\right \vert \right) ^{3}\frac{\hbar kc}{e^{\frac{\hbar kc}{%
K_{B}T}}-1},  \notag \\
&=&8\pi \int_{0}^{1/\hbar \alpha }\frac{k^{2}dk}{\left( 2\pi \right) ^{3}}%
\left( 1-\alpha \hbar k\right) ^{3}\frac{\hbar kc}{e^{\frac{\hbar kc}{K_{B}T}%
}-1},  \notag \\
&=&\int_{0}^{\nu _{\alpha }}d\nu \mathcal{I}_{\alpha }\left( \nu ,\nu
_{\alpha },T,T_{\alpha }\right) ,  \label{black1}
\end{eqnarray}
where
\begin{subequations}
\begin{eqnarray}
\mathcal{I}_{\alpha }\left( \nu ,\nu _{\alpha },T,T_{\alpha }\right)
&=&\left( 1-\frac{\nu }{\nu _{\alpha }}\right) ^{3}\mathcal{I}_{0}\left( \nu
,T\right), \\
\nu _{\alpha }&=&\frac{c}{h\alpha }, \\
T_{\alpha }&=&\frac{c}{K_{B}\alpha },
\end{eqnarray}%
and 
\end{subequations}
\begin{equation}
\mathcal{I}_{0}\left( \nu ,\nu _{\alpha },T,T_{\alpha }\right) =\frac{8\pi }{%
c^{3}}\frac{h\nu ^{3}}{e^{\frac{\nu T_{\alpha }}{\nu _{\alpha }T}}-1}.
\end{equation}%
Here, $\mathcal{I}_{\alpha }\left( \nu ,\nu _{\alpha },T,T_{\alpha }\right)$
and $\mathcal{I}_{0}\left( \nu ,\nu _{\alpha },T,T_{\alpha }\right)$ are the
modified and regular spectral functions, respectively. To observe the effect
of GUP$^{\ast }$ on the shape of the spectral function we plot the functions%
\begin{equation}
\mathcal{F}_{\alpha }\left(\frac{\nu }{\nu _{\alpha }},\frac{T_{\alpha }}{T}%
\right)\mathcal{=}\left( 1-\frac{\nu }{\nu _{\alpha }}\right) ^{3}\mathcal{F}%
_{0},
\end{equation}
and 
\begin{equation}
\mathcal{F}_{0}\left(\frac{\nu }{\nu _{\alpha }},\frac{T_{\alpha }}{T}\right)%
\mathcal{=}\frac{\left( \frac{\nu }{\nu _{\alpha }}\right) ^{3}}{e^{\frac{%
\nu T_{\alpha }}{\nu _{\alpha }T}}-1},
\end{equation}
versus the frequency ratio, $\left({\nu }/{\nu _{\alpha }}\right)\in \left[
0,1\right] $, at three different values of temperatures. When the
deformation parameter is very small the distortion in the spectral functions
is undetectable in low and high frequencies as we present in Fig. \ref{fig5}%
. 

\begin{figure}[hbtp]
\centering
\includegraphics[scale=1]{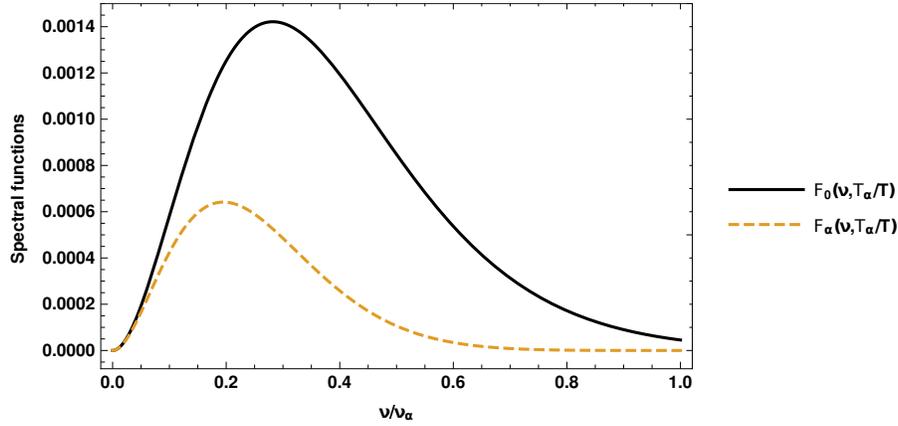}  
\caption{The blackbody radiation spectrum versus frequency in the GUP$^\ast$
framework at temperature $T=0.1T_{\protect \alpha }$.}
\label{fig6}
\end{figure}

For a higher value of deformation parameter spectral functions closely
coincide with each others only in very small frequencies as we demonstrate
in Fig. \ref{fig6}. In such a case the spectral function reaches its maximal
value at a relatively lower frequency value. Moreover, in the GUP$^\ast$
framework, the peak value of the spectral functions, hence, the average
energy is relatively small. 

In Fig \ref{fig6}, we reveal the spectral functions at $T=T_{\alpha }$. We
observe that the deviations increase while the average energy decreases. We
conclude that at high frequencies the blackbody radiations are modified.  
\begin{figure}[hbtp]
\centering
\includegraphics[scale=1]{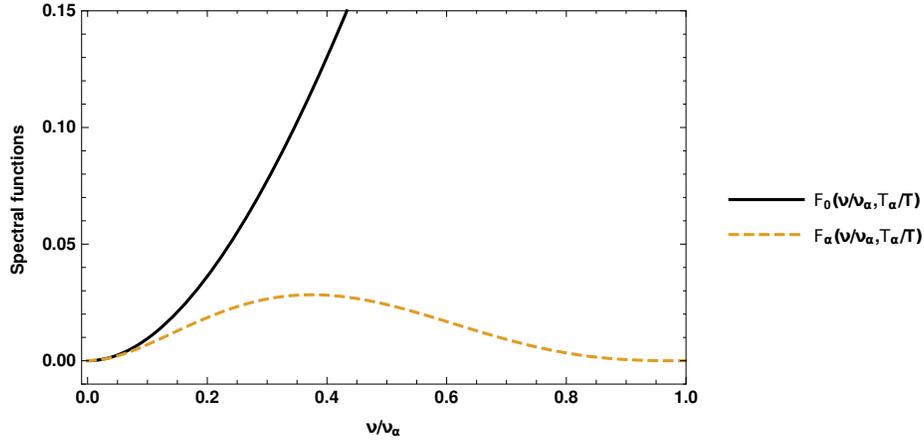}  
\caption{The blackbody radiation spectrum versus frequency in the GUP$^\ast$
framework at temperature }
\label{fig7}
\end{figure}

Our findings approve the studies of Chang \emph{et al.} \cite{Changetal20022}
and Pedram \cite{Pedram2012}. The distortion in the blackbody radiation can
not be observed until the temperature is high enough.





\subsection{The cosmological constant}

Finally, we discuss the cosmological constant in the framework of the GUP$%
^{\ast }$. It is well-known that the cosmological constant can be achieved
by summing over the momentum states of the zero-point energies of the
harmonic oscillator \cite{caroll,Weinberg}. So that, we employ the standard
form of the zero-point energy of each oscillator of mass $m$%
\begin{equation}
\frac{\hbar \omega }{2}=\frac{1}{2}\sqrt{p^{2}+m^{2}},
\end{equation}%
to sum over all momentum states per unit volume. We get%
\begin{eqnarray}
\Lambda \left( m\right) &=&\frac{1}{2}\int d^{3}p\left( 1-\alpha \left \vert
p\right \vert \right) ^{3}\sqrt{p^{2}+m^{2}},  \notag \\
&=&2\pi \int_{0}^{1/\alpha }dp\left( 1-\alpha p\right) ^{3}p^{2}\sqrt{%
p^{2}+m^{2}}.  \label{3A}
\end{eqnarray}
We perform the substitution $k=\frac{p}{m}$, and then we find 
\begin{equation}
\Lambda \left( m\right) =2\pi m^4 \int_{0}^{a}dk\left( 1-\frac{k}{a}\right)
^{3}k^{2}\sqrt{1+k^{2}},
\end{equation}%
where $a=1/\left( \alpha m \right)$. After we take the integrals, we arrive
at%
\begin{eqnarray}
\Lambda \left( m\right) &=&2\pi m^4 \left \{ \left[ \frac{a\left(
a^{2}+1\right) ^{\frac{3}{2}}}{4}-\frac{a\sqrt{a^{2}+1}}{8}-\frac{\mathrm{%
arcsinh}a}{8}\right] -\frac{3}{a} \left[ \frac{2}{15}+\frac{a^{2}\left(
a^{2}+1\right) ^{\frac{3}{2}}}{5}-\frac{2}{15}\left( a^{2}+1\right) ^{\frac{3%
}{2}}\right] \right.  \notag \\
&&\left. +\frac{3}{a^2} \left[ \frac{a^{3}\left( a^{2}+1\right) ^{\frac{3}{2}%
}}{6}-\frac{a\left( a^{2}+1\right) ^{\frac{3}{2}}}{8}+\frac{a\sqrt{a^{2}+1}}{%
16}+\frac{\mathrm{arcsinh}a}{16}\right] \right.  \notag \\
&&\left. -\frac{1}{a^3}\left[ -\frac{8}{105}+\frac{a^{4}\left(
a^{2}+1\right) ^{\frac{3}{2}}}{7}-\frac{4}{35}a^{2}\left( a^{2}+1\right) ^{%
\frac{3}{2}}+\frac{8}{105}\left( a^{2}+1\right) ^{\frac{3}{2}}\right] \right
\}.
\end{eqnarray}%
In the massless case it reduces into the form of%
\begin{equation}
\Lambda \left( 0\right) =\Lambda _{GUP^{\ast }}=\frac{\pi }{70\alpha ^{4}}=%
\frac{\pi}{70}\left( \frac{\hbar }{\ell _{P}}\right) ^{4},  \label{3B}
\end{equation}
where $\ell _{P}$ is the Planck length. Note that due to the GUP$^{\ast }$
influence, the cosmological constant is automatically rendered finite acting
effectively as the UV cutoff.

At this point to end the paper with a comparison, we recall the cosmological
constant predictions that exist in the literature \cite%
{Kempef,Changetal20022, pedram}. 
\begin{eqnarray}
\Lambda _{KMM}&=&\frac{\pi }{2}\left( \frac{\hbar }{\ell _{P}}\right) ^{4},
\label{3C} \\
\Lambda _{Pedram}&=&\frac{\pi 64}{3645}\left( \frac{\hbar }{\ell _{P}}%
\right)^{4}.  \label{3D}
\end{eqnarray}
Then, we compare our finding, Eq. (\ref{3B}), with the others Eqs. (\ref{3C}%
) and (\ref{3D}). We find 
\begin{eqnarray}
\Lambda _{GUP^{\ast }} &<&\Lambda _{Pedram}<\Lambda _{KMM}, \\
\frac{\Lambda _{GUP^{\ast }}}{\Lambda _{Pedram}} &=&0.8; \quad \frac{\Lambda
_{GUP^{\ast }}}{\Lambda _{KMM}}=0.\,028.
\end{eqnarray}
This means that the model proposed by Chung and Hassanabadi is interesting
since it gives the smallest amount among these massless cosmological
constants. Therefore, we conclude that this approximation is more
appropriate than the others.

\section{Conclusion}

In this paper, we used a higher order generalized uncertainty principle that
is presented by Chung and Hassanabadi which predicts a minimal length
uncertainty and a maximal observable momentum value. After a brief
introduction of the new deformed formalism, we studied a one dimensional
classical ideal gas in a canonical ensemble. We constructed the partition
function by taking the deformed commutation relations into account. We
derived the internal and Helmhotz free energy, reduced entropy and specific
heat functions. Then, we have depicted them by employing numerical values.
We discussed the effect of the deformation parameters on the thermal
quantities. We observed that, one can observe these effects only in very
high temperatures. Then, we discussed the blackbody radiation and
cosmological constant problems according to the new approach. We showed that
the latter formalism gave more accurate results after a comparison with the
existing ones in the literature.

\section*{Acknowledgment}

B.C. L\" utf\"uo\u{g}lu, was partially supported by the
Turkish Science and Research Council (T\"{U}B\.{I}TAK).

\end{document}